# Unimode material based low-frequency underwater acoustic isolation

Yu Wei [a], Binghao Zhao [b], Fen Du [a, 1], Yi Chen [c, d] and Gengkai Hu [a]

[a] Marine Science and Technology Domain, Beijing Institute of Technology, Zhuhai 519088, China
[b] School of Chemical Engineering, Dalian University of Technology, Dalian 116024, China
[c] Institute of Nanotechnology, Karlsruhe Institute of Technology (KIT), Karlsruhe 76128, Germany
[d] Institute of Applied Physics, Karlsruhe Institute of Technology (KIT), Karlsruhe 76128, Germany

**Abstract:**

Extremal materials are a specific class of Cauchy materials whose elasticity tensor has one or more zero eigenvalues. Each zero eigenvalue corresponds to a soft mode requiring zero strain energy, while non-zero eigenvalues correspond to hard modes that cost energy. According to the number, $N$, of zero eigenvalues, these materials can be referred to as unimode ($N = 1$), bimode ($N = 2$), etc. Extremal materials have enabled novel functions beyond conventional Cauchy media, e.g., phonon polarizers, Rayleigh wave isolators and underwater acoustic cloaks. These functions typically require a single extremal material. Interfaces between two extremal materials exhibit rich wave behaviors, yet have been seldom explored. Here, we proposed the concept of complementary extremal materials, i.e., the soft mode of one extremal material is a hard mode of the other. As one example, we study the interface between an isotropic unimode material and an isotropic bimode material. We show that the interface allows perfect mode conversion from longitudinal waves to transverse waves. A low-frequency underwater acoustic insulator based on complementary extremal materials is proposed. Our finding has been verified with designed metamaterials and using effective-medium modeling. This work demonstrates the potential of complementary extremal materials in controlling elastic wave polarization and waterborne sound.

---

[1] Corresponding author.
  *Email address:* fendu@bit.edu.cn (Fen Du)

# 1. Introduction

Isolating waterborne sound is crucial in underwater acoustic engineering, with applications including noise reduction for offshore equipment [1-2], shielding of underwater sound radiation [3], enhancement of underwater communications [4] and precise control of underwater acoustics [5-6], etc. Conventional passive sound insulation materials typically rely on impedance mismatch [3, 7-10]. In general, a larger impedance contrast between the insulating material and the background medium results in better sound isolation. Because the impedance of common solids is several orders of magnitude higher than that of air, even thin solid plates can effectively block airborne sound. However, the impedance of water is comparable to that of common solid (e.g., metals), which means that effectively blocking waterborne sound (particularly at low frequencies) requires either thick and rigid solid plates or soft inclusions such as air bubbles [11-12]. The former approach is bulky, whereas the latter lacks structural integrity under pressure, making the development of compact, low-frequency underwater acoustic isolators both challenging and of high practical value.

The existence of elastic metamaterials breathed vitality into classical continuum mechanics. Extremal materials can exhibit quite different wave properties than common materials [13-14], and have recently attracted considerable attention [15-23]. They are mathematically defined as elastic materials whose elasticity tensor can be represented as a $6 \times 6$ matrix using Voigt or Kelvin notation with $N \geq 1$ zero eigenvalues [24]. Thus, extremal materials are a kind of Cauchy medium whose constraints are relaxed, having inherently soft deformation modes that, in principle, cost no energy. Depending on the value of $N$, we refer them as unimode materials ($N = 1$, UM), bimode materials ($N = 2$, BM), trimode materials ($N = 3$, TM), quadramode materials ($N = 4$, QM) and pentamode materials ($N = 5$, PM), following the nomenclature of a previous paper [24]. These materials can be well constructed based on squares or beams connected by extremely thin parts [25-29]. Although the eigenvalues are typically not precisely zero, they can be engineered to be orders of magnitude smaller than others [26]. Each of the zero eigenvalue is associated to a soft deformation mode of the material in the sense that the deformation requires conceptually zero energy. Conventional Cauchy continuum corresponds to $N = 0$ with no soft mode. The presence of soft modes in extremal materials can drastically change their bulk wave properties [13-14]. There are many studies reporting that complex acoustic/elastic wave path control and polarization modulation can be achieved with an appropriate single type of extremal elastic material. For example, underwater acoustic cloaking [22, 30-31], waterborne sound metasurface [23, 28], hydraulic silencers [32] and load-bearing waterborne sound insulation [8] can be realized by carefully designed PM. The QM can be capable of blocking mode conversion at solid-liquid interfaces [33], achieving shear wave phonon polarizers [29] and tunable elastic wave quarter-wave plate [34]. However, these exotic wave functions all rely on a single type of extremal material, and there are few reports on the wave properties of the combination of extremal material. Such as the propagation characteristics of elastic waves at an interface consisting of isotropic PM and isotropic UM, which may be of great value in low-frequency underwater acoustic wave control.

When a soft mode of one extremal material coincides with a hard mode of another extremal material, we refer these two extremal materials as complementary extremal materials. In this paper, we aim to explore the propagation properties of elastic waves at interfaces between 2D complementary extremal materials. It is found in this paper that the complementary extremal material of water enables the design of low-frequency flow-permeable underwater acoustic isolator. Unlike methods that employ the low impedance arising from the extreme anisotropy of PM to block waterborne sound [7-8, 32], the underwater acoustic isolator designed here is realized through mode conversion at the interface of isotropic complementary extremal materials. Subsequently, we designed isotropic complementary extremal materials and verified our findings based on continuum theory with the metamaterial model. It should be noticed that 2D BM are also called PM in previous literature [13], since 2D BM have the same static and dynamic characteristics as 3D PM in the plane.

The paper is organized as follows: the wave characteristics and microstructure design of complementary extremal materials are detailed in Section 2. The effective elasticity constants of truss model are given analytically. In section 3, the wave properties at the interface of isotropic complementary extremal materials are analyzed and validated using both truss models and their corresponding solid models. As an application, a low-frequency flow-permeable underwater acoustic isolator based on block array made of isotropic UM is proposed and demonstrated in Section 4. In the end, the concluding section summarized the main findings of the work.

## 2. Complimentary extremal materials

This section begins by introducing the concept of complementary extremal materials, followed by an analysis of elastic wave propagation characteristics at their interface, with a focus on the two-dimensional (2D) isotropic case for simplicity.

*2.1 Wave characteristics*

The fourth-order elasticity tensor, $C_{ijkl}$, for an elastic material with $N$ zero eigenvalues can be expressed in form of Kelvin's decomposition as $C_{ijkl} = \sum_{r=1}^{3-N} K_r S_{ij}^{(r)} S_{kl}^{(r)}$ [24], with $K_r$ being the non-zero eigenvalues of the elasticity tensor and $\mathbf{S}^{(r)}$ being a second order symmetric tensor (characteristic tensor, hard mode), which implying that this material is stiff to any stress in the subspace spanned by $\mathbf{S}^{(r)}$. For convenience, in the following derivation, the elasticity tensor can also be written as a compact form

$$C_{ijkl} = \sum_{r=1}^{3-N} S_{ij}^{(r)} S_{kl}^{(r)}, \tag{1}$$

where $K_r$ is absorbed in $\mathbf{S}^{(r)}$. Then, the equations of motion in the absence of body force and the

constitutive law of a Cauchy elastic material are

$$\sigma_{ij,j} = \rho \frac{\partial^2 u_i}{\partial t^2}, \sigma_{ij} = \frac{1}{2} C_{ijkl}\left(u_{k,j} + u_{j,k}\right), \tag{2}$$

in which, $\sigma_{ij}$ is the symmetric stress tensor, $u_i$ denotes the displacement, $t$ is time, $\rho$ denotes the mass density. Repeated indices should be understood as Einstein summation. The comma denotes spatial derivative. Then, for a time-harmonic elastic plane wave, the Christoffel's equation is given by [35]

$$\left(\Gamma_{ik} - \rho c^2 \delta_{ik}\right)\hat{u}_k = 0, \ \Gamma_{ik} = C_{ijkl} n_j n_l, \tag{3}$$

where $c = \omega/k$ is the phase velocity, $\omega$ and $k$ the angular frequency and wavenumber, respectively. $\boldsymbol{\Gamma}$ is the acoustic tensor, $\hat{\mathbf{u}}$ the particles polarization, $\mathbf{n}$ the direction cosine of the wave propagation, $\delta_{ik}$ the Kronecker delta. To obtain a nontrivial solution, the determinant of Eq. (3) should be vanishing

$$\left|\Gamma_{ik} - \rho c^2 \delta_{ik}\right| = 0. \tag{4}$$

By combining Eq. (1) and Eq. (4), the dispersion relation for a general 2D extremal elastic material can be calculated. The corresponding polarization $\hat{\mathbf{u}}$ is subsequently obtained by substituting the calculated phase velocity $c$ back into Eq. (3). Generally, for an anisotropic 2D extremal elastic material, the closed formula of Eq. (4) cannot be obtained and numerical algorithm is needed to derive the solution. As we pursue closed-form formula, only the orthotropic extremal elastic material is considered in this paper.

In 2D case, there are two types of extremal materials, i.e., unimode material ($N = 1$, UM, which possess a unique soft mode $\mathbf{E}_0$) and bimode material ($N = 2$, BM, which possess a unique hard mode $\mathbf{S}_1$) [24]. Specifically, when $\mathbf{E}_0 = \xi \mathbf{S}_1 \ (\xi \neq 0)$, these two extremal materials are referred to as complementary extremal materials, which resulting in the stress fields in the complementary extremal materials being always orthogonal to each other, i.e., $\boldsymbol{\sigma}^{\text{BM}}:\boldsymbol{\sigma}^{\text{UM}} = 0$. For instance, water behaves as an isotropic BM characterized by a Poisson's ratio of $+1$ and a hard mode corresponding to hydrostatic stress [13]. In contrast, a 2D isotropic UM exhibits a Poisson's ratio of $-1$ and a soft mode corresponding to hydrostatic mode. These two extremal materials are thus complementary, i.e., the hard mode of water coincides with the soft mode of the isotropic UM. Notably, the complementary extremal material of water may not be unique, since it only imposes requirements on the soft mode of UM without any restrictions on its hard mode. Besides, the concept of complementary extremal elastic material can also be generalized to 3D case.

For an orthotropic UM, the elasticity matrix in the principal coordinates system is then, in Voigt's notation, can be expressed as the following two types of $3 \times 3$ matrices

$$\mathbf{C} = \begin{bmatrix} C_{11} & \alpha\sqrt{C_{11}C_{22}} & 0 \\ \alpha\sqrt{C_{11}C_{22}} & C_{22} & 0 \\ 0 & 0 & C_{33} \end{bmatrix}, \ |\alpha| = 1, \ C_{33} \neq 0. \tag{5}$$

$$\mathbf{C} = \begin{bmatrix} C_{11} & \alpha\sqrt{C_{11}C_{22}} & 0 \\ \alpha\sqrt{C_{11}C_{22}} & C_{22} & 0 \\ 0 & 0 & 0 \end{bmatrix}, \ |\alpha| < 1, \tag{6}$$

The semi-positive definiteness of elasticity tensor requires that $\{C_{11}, C_{22}, C_{33}\} > 0$.

For the UM with the elasticity matrix of Eq. (5), its soft mode is $\mathbf{E}_0 = \left[-\alpha\sqrt{C_{22}},\ \sqrt{C_{11}},\ 0\right]^\mathrm{T}$, in which $\alpha = +1$ implies it cannot support a stress state that is tensile in one principal direction while compressive in the other, whereas $\alpha = -1$ indicates it cannot support stress state that are simultaneously tensile or compressive in both principal directions. In this case, the elasticity matrix of BM complementary to UM with Eq. (5) is as follows,

$$\mathbf{C} = \begin{bmatrix} C_{22} & -\alpha\sqrt{C_{11}C_{22}} & 0 \\ -\alpha\sqrt{C_{11}C_{22}} & C_{11} & 0 \\ 0 & 0 & 0 \end{bmatrix}, \ |\alpha| = 1. \tag{7}$$

The stress in BM is always proportional to its hard mode $\mathbf{S}_1 = \left[-\alpha\sqrt{C_{22}},\ \sqrt{C_{11}},\ 0\right]^\mathrm{T}$, i.e., $\mathbf{S}_1 = \mathbf{E}_0$. According to Eq. (7), the principal directions of the BM coincide with those of its characteristic tensor. The cases of $\alpha = +1$ and $\alpha = -1$ correspond to negative and positive BM (which are also referenced as negative and positive PM), exhibiting negative and positive Poisson's ratios, respectively. Furthermore, the BM with $\alpha = +1$ can only support stress where on principal direction is under tension and the other under compression. In contrast, when $\alpha = -1$, it can only support stress that are either simultaneously tensile or compressive along its principal directions. Specifically, when $\alpha = -1$ and $C_{11} = C_{22}$, it becomes an isotropic BM, in which case the hydrostatic stress is its hard mode.

Substituting the UM elasticity matrix of Eq. (5) into Eq. (4) yields the dispersion relation

$$\begin{aligned}
\omega_1^2 &= \frac{1}{2\rho}\left(k^2(C_{11}+C_{33}) + k_2^2(C_{22}-C_{11}) - \sqrt{k_2^2 C_n + k^4(C_{11}-C_{33})^2}\right), \\
\omega_2^2 &= \frac{1}{2\rho}\left(k^2(C_{11}+C_{33}) + k_2^2(C_{22}-C_{11}) + \sqrt{k_2^2 C_n + k^4(C_{11}-C_{33})^2}\right), \\
C_n &= k_2^2\left((C_{11}-C_{22})^2 - 4C_{33}(C_{11}+C_{22})\right) \\
&\quad + 2k^2\left(C_{11}C_{22} + C_{22}C_{33} + 3C_{11}C_{33} - C_{11}^2\right) \\
&\quad + 8\alpha k_1^2 C_{33}\sqrt{C_{11}C_{22}},
\end{aligned} \tag{8}$$

The two bulk wave modes corresponding to Eq. (8) are as follows

$$\hat{\mathbf{u}}_1 = \begin{bmatrix} C_{33} - C_{11} + (C_{11} + C_{22})n_2^2 - 2C_{33}n_2^2 + \sqrt{n_2^2 C_n + (C_{11} - C_{33})^2} \\ -2n_1 n_2 \left(C_{33} + \alpha\sqrt{C_{11}C_{22}}\right) \end{bmatrix},$$

$$\hat{\mathbf{u}}_2 = \begin{bmatrix} C_{33} - C_{11} + (C_{11} + C_{22})n_2^2 - 2C_{33}n_2^2 - \sqrt{n_2^2 C_n + (C_{11} - C_{33})^2} \\ -2n_1 n_2 \left(C_{33} + \alpha\sqrt{C_{11}C_{22}}\right) \end{bmatrix}.$$

(9)

According to Eqns. (8)-(9), the polarizations and equal frequency curves (EFCs) shapes of the bulk wave are dependent on both elastic constants ($C_{11}$, $C_{22}$ and $C_{33}$) and the sign of $\alpha$. For example, when $\alpha = +1$ the outer EFC has four openings and the inner EFC is closed, while $\alpha = -1$ leading to two closed EFCs (see detail in Appendix). In particular, when $\alpha = -1$ and $C_{11} = C_{22} = C_{12} = K$ ($K > 0$), the UM is isotropic, and its acoustic tensor degenerates to $\mathbf{\Gamma} = K\mathbf{I}$, in which $\mathbf{I}$ is identity tensor. At this point, the EFC of the isotropic UM are overlapped circles (i.e., they possess the same wave velocity), and its bulk waves can admit arbitrary polarizations. This feature will provide new possibilities for polarization control of elastic waves.

Furthermore, substituting the elasticity matrix of Eq. (7) (i.e., BM complemented to the UM of Eq.(5)) into Eq. (4) yields the dispersion relation

$$\omega_1^2 = 0, \quad \omega_2^2 = \left(C_{22}k_1^2 + C_{11}k_2^2\right)/\rho, \tag{10}$$

and the two bulk wave modes corresponding to Eq. (10) are as follows

$$\hat{\mathbf{u}}_1 = \begin{bmatrix} n_2\sqrt{C_{11}} \\ \alpha n_1 \sqrt{C_{22}} \end{bmatrix}, \hat{\mathbf{u}}_2 = \begin{bmatrix} -n_1\sqrt{C_{22}} \\ \alpha n_2 \sqrt{C_{11}} \end{bmatrix}. \tag{11}$$

In this case, the remaining EFC is a closed ellipse whose shape is independent of the sign of $\alpha$, while the polarizations are closely related to the sign of $\alpha$. When the constraints of $\alpha = -1$ and $C_{11} = C_{22} = K$ ($K > 0$) holds, the BM is isotropic (e.g., water, complementary to isotropic UM), and its acoustic tensor is $\mathbf{\Gamma} = K\mathbf{nn}^{\mathrm{T}}$, resulting in the eigenvector corresponding to its non-zero eigenvalue being $\hat{\mathbf{u}}_2 = \mathbf{n}$, i.e., it can only support the longitudinal wave propagation.

As the subsequent discussion mainly focuses on the isotropic case, a brief analysis of the EFC and polarization characteristics of the anisotropic case of the above-mentioned UM, as well as those of the UM with Eq. (6) and its complementary BM, is provided in the Appendix.

*2.2 Microstructure design*

The periodically pin-joint cubic symmetry truss, capable of realizing isotropic UM and isotropic BM, are shown in Fig. 1(a) and (d), respectively, while their corresponding unit cells are illustrated in Fig. 1(b) and 1(e). The out-of-plane thickness of rods are set to be unity unless explicitly specified otherwise. In the following, their effective properties and bulk wave characteristics will be detailed.

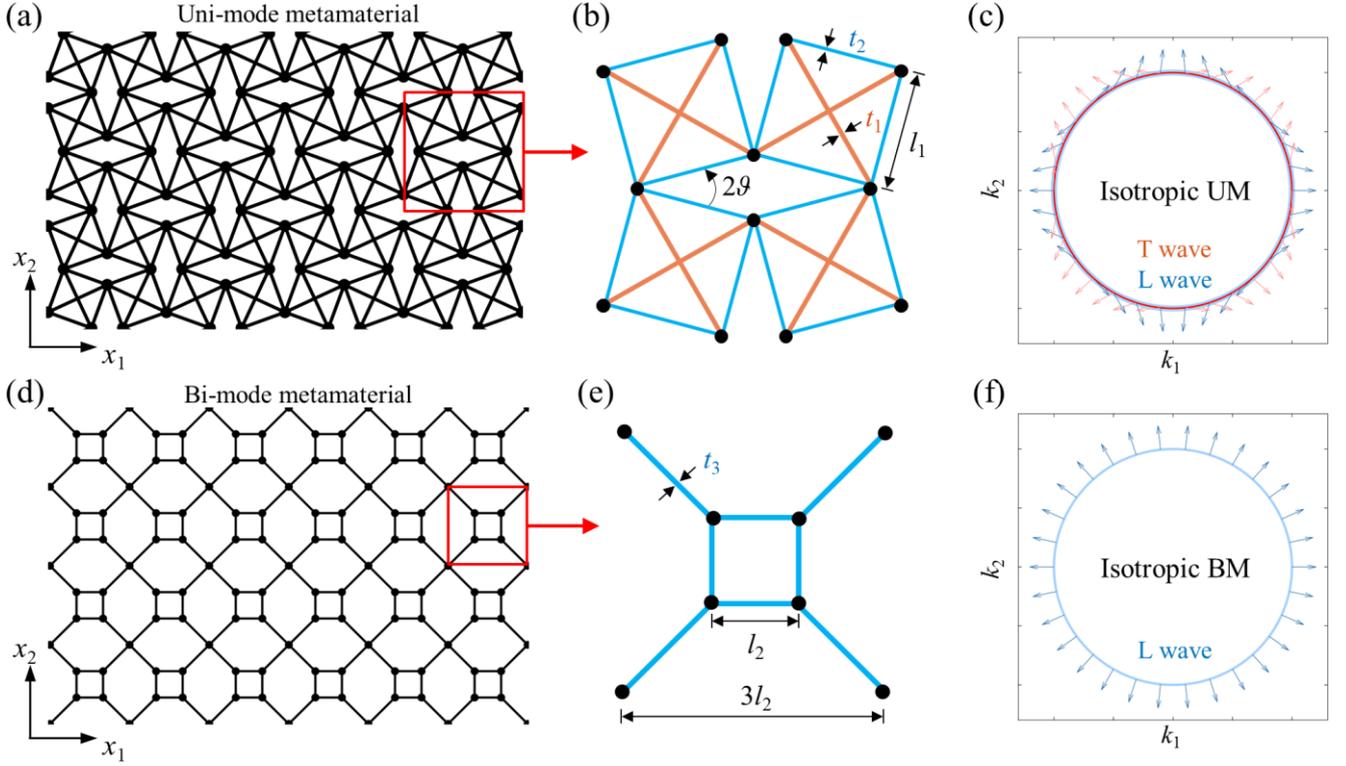

Fig. 1. Sketch of UM and BM truss model and their equi-frequency curves (EFCs). (a) UM metamaterial. (b) The unit cell of UM, which is composed of four small squares with twist angle $\vartheta \in (0, \pi/4)$, containing 12 nodes and 24 rods. The red bars connect the diagonally opposite nodes of the square, while the blue bars link the nodes along its perpendicular edges. The in-plane thickness of the red and blue rods in squares are $t_1$ and $t_2$, respectively. The length of blue rods is $l_1$. (c) The EFCs of the UM in panel(b) when $t_1/t_2 = \sqrt{2}$. The arrows represent the polarizations of the EFCs, shown in the corresponding colors. They are obtained by averaging the displacements of all independent nodes within the unit cell, confirming the structure behaves as an ideal isotropic UM. (d)-(f) same as (a)-(c) but the UM is replaced by BM. The unit cell contains 8 nodes and 8 rods in total. All the rods have the same in-plane thickness $t_3$. The side length of the inside square is $l_2$.

Firstly, we consider the microstructure design sketched in Fig. 1(a) and (b), i.e., a kind of rotating square lattice. Any neighboring unit cell may be reached by translating the reference unit cell with a direct lattice translation vector $x_j \mathbf{a}_j$, with $j \in \{1, 2\}$, where $x_j$ is any set of integer values and $\mathbf{a}_j$ is the lattice vector:

$$\mathbf{a}_1^{(\text{UM})} = 2l_1 (\cos\vartheta + \sin\vartheta)[1,\ 0]^{\text{T}}, \quad \mathbf{a}_2^{(\text{UM})} = 2l_1 (\cos\vartheta + \sin\vartheta)[0,\ 1]^{\text{T}}. \tag{12}$$

Following the Cauchy–Born hypothesis-based homogenization method [34, 36], when $t_1/t_2 = \sqrt{2}$, this lattice can be homogenized as an isotropic UM with elasticity matrix

$$\mathbf{C}^{\text{UM}} = \frac{E_s t_2}{2l_1} \begin{bmatrix} 1 & -1 & 0 \\ -1 & 1 & 0 \\ 0 & 0 & 1 \end{bmatrix}, \tag{13}$$

in which, $E_s$ is the Young's modulus of the rods. Since we consider a frequency range far below the

resonance frequency, we obtain the effective mass density $\rho^{\text{UM}}$ from mass average

$$\rho^{\text{UM}} = \frac{8t_2\rho_1}{l_1(1+\sin 2\vartheta)}, \tag{14}$$

where $\rho_1$ represents the mass density of the rods in UM lattice. It can be seen that, when $t_1/t_2 = \sqrt{2}$, the effective elasticity matrix of such rotating square lattice is independent of the twist angle $\vartheta$, i.e., the $\mathbf{C}^{\text{UM}}$ does not vary with the $\vartheta$. However, the effective density $\rho^{\text{UM}}$ decreases monotonically with the increase of $\vartheta$. Therefore, the bulk waves phase velocities of this isotropic UM increase monotonically with the increases of $\vartheta$. This characteristic makes it easier to realize a perfect interface between the isotropic complementary extremal materials using microstructures.

Furthermore, the wave characteristics of the periodic pin-jointed structure can be evaluated by directly solving the dynamical matrix [37], and its eigenvalues and eigenvectors correspond to the natural frequencies and normal modes, respectively. To obtain the dynamical matrix, the Bloch's theorem is applied to the unit cell [33, 36]. When $t_1/t_2 = \sqrt{2}$, the equifrequency curves (EFCs) of the constructed discrete UM lattice as shown in Fig. 1(c) is exactly the same as the theoretically predicted shapes, i.e., two overlapped circles. In this case, it is an ideal isotropic UM, capable of supporting bulk wave with arbitrary polarization. To ensure consistency with the solid model results (see section 3.3), the polarizations corresponding to the limit $t_1/t_2 \to \sqrt{2}$ are presented, which also satisfy Eq. (9), indicating the evolution from orthotropy to isotropy.

Next, consider the periodic truss model sketched in Fig. 1(d) and (e). For the unit cell shown in Fig. 3(e), one gets

$$\mathbf{a}_1^{(\text{BM})} = 3l_2[1,\ 0]^{\text{T}},\ \mathbf{a}_2^{(\text{BM})} = 3l_2[0,\ 1]^{\text{T}}. \tag{15}$$

Following the Cauchy–Born hypothesis-based homogenization method, the macroscopic elasticity matrix of this lattice can be written as

$$\mathbf{C}^{\text{BM}} = \frac{(2\sqrt{2}-1)E_s t_3}{7l_2}\begin{bmatrix} 1 & 1 & 0 \\ 1 & 1 & 0 \\ 0 & 0 & 0 \end{bmatrix}, \tag{16}$$

i.e., an isotropic BM. In addition, the effective mass density $\rho^{\text{BM}}$ is

$$\rho^{\text{BM}} = \frac{4(\sqrt{2}+1)t_3\rho_2}{9l_2}, \tag{17}$$

where $\rho_2$ denotes the mass density of the rods in BM lattice. The EFC and polarization of the constructed discrete BM lattice as shown in Fig. 1(f) is exactly the same as those predicted by theorem, i.e., it can only support the longitudinal wave propagation.

Then, a question naturally arises: how do the elastic waves propagate at the interface between two

semi-infinite isotropic complementary extremal materials? We notice that, when

$$\tan \vartheta = 0.5, \quad l_1 = \frac{\sqrt{5}}{2} l_2, \tag{18}$$

The two lowest nodes of the UM unit cell will coincident with the two vertices of the central square of the BM unit cell, leading to a perfectly connection between the two lattices (see the zoomed plot in figure 2(c)). In the next section, we focus on the propagation characteristics of elastic waves at the interface formed by isotropic 2D UM and its complementary BM.

## 3. Reflection and refraction

In this section, the reflection and refraction factors are derived for a perfect interface between two semi-infinite isotropic complementary extremal materials, and verified by both continuous and discrete models.

*3.1 Transmission characteristics*

A conventional Cauchy elasticity possesses two bulk waves in 2D case, namely longitudinal wave (short as L wave) and transverse wave (short as S wave) [35]. When a plane wave (e.g., L wave) of amplitude $A_{IL}$ is incident with angle $\theta_{IL}$ onto the solid-solid interface between two half spaces, there are usually reflected waves of amplitudes $A_{RL}$ and $A_{RS}$ and angles $\theta_{RL}$ and $\theta_{RS}$ for L and S waves, respectively. The interface is at $x_2 = 0$. In addition, the transmitted waves have amplitudes $A_{TL}$ and $A_{TS}$ and angles $\theta_{TL}$ and $\theta_{TS}$.

According to the analysis in the previous section, for the case where a plane wave is incident onto the interface between two semi-infinite isotropic complementary extremal materials (lower half space for BM and upper half space for UM), as shown in Fig. 2(a), the reflected wave $A_{RS}$ will disappear. Furthermore, since isotropic UM supports bulk waves with arbitrary polarizations, without loss of generality, the amplitudes of the transmitted waves can be denoted as $A_{T1}$ and $A_{T2}$. Therefore, when a L plane wave of amplitude $A_{IL}$ is incident with angle $\theta_{IL}$ onto the interface from the lower half space, the displacements of the incident, reflected and refracted waves are given by

$$\begin{aligned}
\mathbf{u}_{IL} &= A_{IL} \left[\sin \theta_{IL}, \cos \theta_{IL}\right]^T \exp\left(ik_{IL}\left(x_1 \sin \theta_{IL} + x_2 \cos \theta_{IL}\right) - \omega t\right), \\
\mathbf{u}_{RL} &= A_{RL} \left[\sin \theta_{RL}, -\cos \theta_{RL}\right]^T \exp\left(ik_{RL}\left(x_1 \sin \theta_{RL} - x_2 \cos \theta_{RL}\right) - \omega t\right), \\
\mathbf{u}_{T1} &= A_{T1} \left[\cos \theta_{x1}, \sin \theta_{x1}\right]^T \exp\left(ik_{T1}\left(x_1 \sin \theta_{T1} + x_2 \cos \theta_{T1}\right) - \omega t\right), \\
\mathbf{u}_{T2} &= A_{T2} \left[-\sin \theta_{x1}, \cos \theta_{x1}\right]^T \exp\left(ik_{T2}\left(x_1 \sin \theta_{T2} + x_2 \cos \theta_{T2}\right) - \omega t\right).
\end{aligned} \tag{19}$$

The superscript T means transpose, $\theta_{x1} \in [0, \pi/2]$ represents the angle between any prescribed polarization and the $x_1$-axis. Since one bulk mode of BM vanishing, there would be a discontinuity of particle displacement at the interface. Benefiting from the discontinuity of the particle transverse displacement at the interface, perfect mode conversion between longitudinal wave and transverse wave is expected to be achieved.

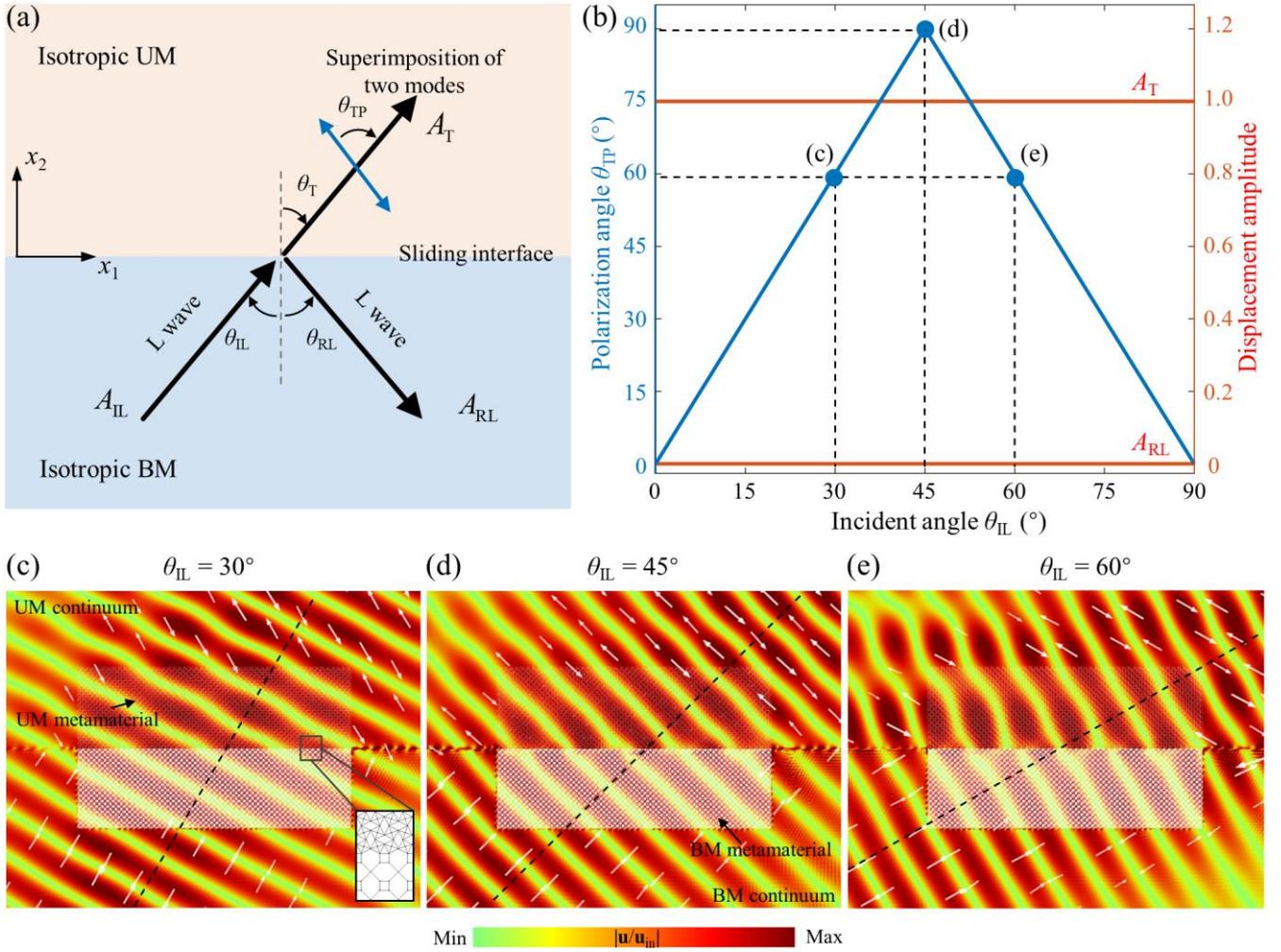

Fig. 2. Theoretical analysis and numerical validation of wave transmission at the interface between two half spaces of isotropic BM and isotropic UM. (a) Diagram of a L wave obliquely incident onto the interface. (b) The polarization angle $\theta_{TP}$, displacement amplitude $A_{RL}$ and $A_T$ versus the incident angle $\theta_{IL}$. (c)-(e) show the calculated displacement amplitude field for the case shown in panel (a) at $\theta_{IL} = \pi/6$, $\pi/4$ and $\pi/3$, respectively. The model is composed of the isotropic complementary extremal lattice (the corresponding microstructures are shown in Fig. 1) and their counterpart. Each type of lattice consists of $N_x \times N_y = 50 \times 15$ unit cells, containing a total of 24000 rods. The zoomed plot shows the detail at the interface. The L wave Gaussian beam with frequency $f = 3.6$ kHz is oblique incident from the BM effective homogeneous counterpart onto the BM-UM lattice, where perfect wave transmission is observed. The PMLs are set to be invisible. The black dashed line and white arrows represent the wave propagation direction and displacement vectors, respectively, indicating the polarization in UM is always symmetrical to the wave propagation direction with respect to the $x_2$-axis.

The boundary conditions for a perfect interface between complementary extremal materials are continuity of transmission energy (it can be also regard as continuity of particle displacement along the characteristic force vector $\mathbf{t} = \mathbf{S}_1 \cdot \mathbf{n}$ of BM [14]) and continuity of stress at the interface [27], which can be expressed as

$$\mathbf{u}^{\mathrm{I}} \cdot \left( \mathbf{S}_1 \cdot \mathbf{n}^{\perp} \right) = \mathbf{u}^{\mathrm{II}} \cdot \left( \mathbf{S}_1 \cdot \mathbf{n}^{\perp} \right),$$
$$\boldsymbol{\sigma}^{\mathrm{I}} \cdot \mathbf{n}^{\perp} = \boldsymbol{\sigma}^{\mathrm{II}} \cdot \mathbf{n}^{\perp},$$
(20)

in which, $\mathbf{n}^{\perp} = [0, 1]^{\mathrm{T}}$ is the normal of the interface. The superscript I and II correspond to the background medium in the lower half space and the transmission medium in the upper half space, respectively.

Besides, Snell's law states that

$$k_{\mathrm{IL}} \sin \theta_{\mathrm{IL}} = k_{\mathrm{RL}} \sin \theta_{\mathrm{RL}} = k_{\mathrm{T1}} \sin \theta_{\mathrm{T1}} = k_{\mathrm{T2}} \sin \theta_{\mathrm{T2}}.$$
(21)

Let's consider the isotropic UM and isotropic BM have the same magnitude of elastic constants and density, i.e., $C_{11}^{\mathrm{UM}} = C_{11}^{\mathrm{BM}} = K$ and $\rho^{\mathrm{UM}} = \rho^{\mathrm{BM}} = \rho$, which leading to the UM and BM share the same wave velocity $c_{\mathrm{L}} = c_{\mathrm{T1}} = c_{\mathrm{T2}} = \sqrt{K/\rho}$. Then, according to Eq. (21), we have $\theta_{\mathrm{IL}} = \theta_{\mathrm{RL}} = \theta_{\mathrm{T1}} = \theta_{\mathrm{T2}} = \theta$ and $k_{\mathrm{IL}} = k_{\mathrm{RL}} = k_{\mathrm{T1}} = k_{\mathrm{T1}} = k$. Subsequently, combing Eq. (2) and Eqns. (19)-(21) yields three homogeneous equations

$$\begin{bmatrix} \cos\theta & \sin\theta_x & \cos\theta_x \\ 0 & \cos(\theta-\theta_x) & \sin(\theta-\theta_x) \\ 1 & \sin(\theta-\theta_x) & -\cos(\theta-\theta_x) \end{bmatrix} \begin{bmatrix} A_{\mathrm{RL}} \\ A_{\mathrm{T1}} \\ A_{\mathrm{T2}} \end{bmatrix} = A_{\mathrm{IL}} \begin{bmatrix} \cos\theta \\ 0 \\ -1 \end{bmatrix}.$$
(22)

Without loss of generality, let $A_{\mathrm{IL}} = 1$. Then, the unknowns of amplitudes $A_{\mathrm{RL}}, A_{\mathrm{T1}}$ and $A_{\mathrm{T2}}$ can be calculated by solving Eq. (22)

$$A_{\mathrm{RL}} = 0, \ A_{\mathrm{T1}} = -\sin(\theta - \theta_{x1}), \ A_{\mathrm{T2}} = \cos(\theta - \theta_{x2}).$$
(23)

As a result, the reflected L wave is absent, and the displacement field of the transmitted wave can be superimposed in a compact form as

$$\mathbf{u}_{\mathrm{T}} = [-\sin\theta, \cos\theta]^{\mathrm{T}} \exp\left( \mathrm{i} k (x_1 \sin\theta + x_2 \cos\theta) - \omega t \right),$$
(24)

which indicates the polarization of total displacement field of the transmitted wave is independent of the choice of bulk wave polarization. And the polarization angle $\theta_{\mathrm{TP}} = \min\{\langle \mathbf{u}_{\mathrm{T}}, \mathbf{n} \rangle, \langle \mathbf{u}_{\mathrm{T}}, -\mathbf{n} \rangle\}$, where $\langle \cdot, \cdot \rangle$ denotes the angle between the two vectors. The $A_{\mathrm{RL}}, A_{\mathrm{T}}$ and $\theta_{\mathrm{TP}}$ versus incident angle $\theta_{\mathrm{IL}}$ is shown in Fig. 2(b). It is worth noting that $A_{\mathrm{RL}}$ and $A_{\mathrm{T}}$ are independent of $\theta_{\mathrm{IL}}$, and always equal to 0 and 1, respectively. Besides, $\theta_{\mathrm{TP}}$ is linearly dependent on $\theta_{\mathrm{IL}}$, and when $\theta_{\mathrm{IL}} = \pi/4$, $\theta_{\mathrm{TP}}$ reach its highest value of $\pi/2$, i.e., when the L wave is incident onto this interface at an angle of $\theta_{\mathrm{IL}} = \pi/4$, perfect mode conversion will occur. On the other hand, it can be seen from Eq. (24) that $\mathbf{u}_{\mathrm{T}}$ is always symmetrical with respect to $\mathbf{n}$ about the $x_2$-axis.

This wave function can be also regarded as an elastic wave diode. In other words, at the incident angle corresponding to mode conversion, L wave can propagate across the interface only from the isotropic BM side into the isotropic UM, whereas the L wave propagate across the interface from the isotropic UM side into the isotropic BM is forbidden. Conversely, S wave can pass through the interface only from the isotropic UM into the isotropic BM along the mode-conversion path, but the reverse transmission is not allowed. We remark that this process does not violate reciprocity principle.

### 3.2 Numerical validation with truss model

We verified the above findings by numerical simulations using the commercial finite element analysis (FEA) software COMSOL Multiphysics 5.6, examining both the homogeneous and lattice model. Here, we consider a simulation model as shown in Fig. 2(c). In the simulations, the computed regions are surrounded by the perfectly matched layers (PMLs) to avoid reflectance due to the finite boundary. Since the homogeneous complementary extremal materials are a kind of linear Cauchy elasticity, the PMLs still work.

When the geometric parameters of isotropic UM and BM lattices take the relation of Eq. (18), they will have the same lattice constants: $\mathbf{a}_1 = \mathbf{a}_1^{(UM)} = \mathbf{a}_1^{(BM)}$ and $\mathbf{a}_2 = \mathbf{a}_2^{(UM)} = \mathbf{a}_2^{(BM)}$. Furthermore, according to Eqns. (13)-(14) and Eqns. (16)-(17), when

$$t_2 = \frac{\sqrt{5}(2\sqrt{2}-1)}{7} t_3, \quad \rho_1 = \frac{5+3\sqrt{2}}{20} \rho_2, \tag{25}$$

the effective elasticity matrices and effective mass density of the complementary extremal lattice satisfy the relationships shown in Section 3.1, i.e., $C_{11}^{UM} = C_{11}^{BM}$ and $\rho^{UM} = \rho^{BM}$. The material constants and geometric parameters in simulations are given unless explicitly specified otherwise: $l_2 = 10$ mm, $E_s = 8.6140$ GPa, $t_3 = 0.01$ m and $\rho_2 = 931.98$ kg/m$^3$. For UM lattice, its material constants and geometric parameters are determined by Eqns. (18) and (25). As a results, the effective elastic constants and effective mass densities of the two lattices are $C_{11}^{UM} = C_{11}^{BM} = 2.25$ GPa and $\rho^{UM} = \rho^{BM} = 1000$ kg/m$^3$, and the lattice constants of the two lattices are both $|\mathbf{a}_1| = |\mathbf{a}_2| = a = 30$ mm. The microstructure part in Fig. 2 is composed of isotropic UM/BM lattices. The total size of the microstructure part is $1500$ mm $\times$ $900$ mm.

In order to check our designed isotropic complementary extremal elastic lattices and their wave function, we examine the wave transmission for a Gaussian amplitude modulated L wave beam with frequency $f = 3.6$ kHz incident at angles of $\pi/6$, $\pi/4$ and $\pi/3$ onto the composite interface. The simulated amplitude of displacement field (normalized by the incident wave) are plotted in Fig. 2(c)-(e), respectively, which are conducted by using the Solid Mechanics and Truss Module of COMSOL Multiphysics in the frequency domain. As shown in Figs. 2(c)-(e), perfect transmissions are observed between the isotropic UM/BM lattice and its counterpart, which in fact confirms our design method for the complementary extremal materials. And the simulation results are in good agreement with the theoretical predictions, which not only validate the effective properties of the complementary extremal microstructures, but also the mode conversion function of their interface.

### 3.3 Numerical validation with solid model

When fabrication is considered, these ideal joints between individual beams in the above lattices should be replaced with solid connections. The local geometry of beams should also be modified accordingly to dramatically reduce their bending stiffness, which is known as compliant mechanism [38]. We call in the following solid model for complementary extremal materials made by compliant mechanism. The solid model of the complementary extremal lattice can be realized by tailored 2D lattice composed of double-triangle elements, at which point the triangle tips merge into circles with small (ideally zero) diameter $d$ connecting the triangles. Based on the truss model that shown in Fig. 1(b) and (e), the solid model shown in Fig. 3(a) and (d) are designed, respectively. In order to obtain geometrically well-defined objects, we use the union of circles and several triangle tips. Besides, to ensure that both lattices have the same equivalent mass density, weights are added to the legs of the central square of the BM lattice, as shown in panel (d).

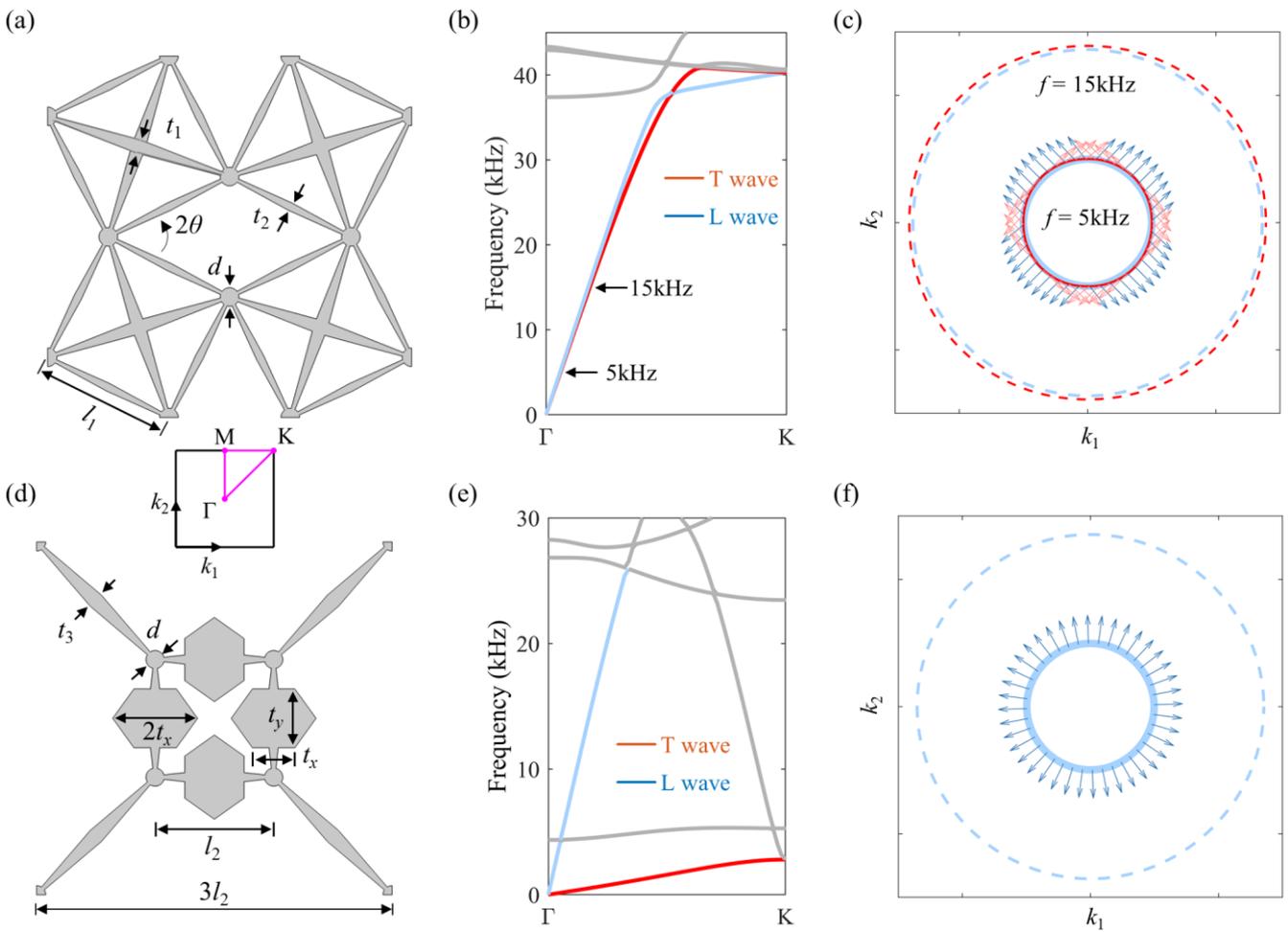

Fig. 3. The solid model of complementary extremal materials and their wave properties. (a) The isotropic UM. (b) and (c) are the band structures in the ΓK direction and the EFCs at frequencies $f = 5$ kHz (solid lines) and $15$ kHz (dashed lines), respectively, for the solid model in the panel (a). The blue band correspond to L wave, and the red to the T wave. The arrows represent polarizations of the modes with corresponding colors, as determined by volume averaging. The first Brillouin zone (BZ) in the reciprocal space is also sketched, where its bounds are depicted as a square. The paths are highlighted in pink. Calculations are performed for infinite samples. (d)-(f) same as (a)-(c) but the UM is replaced by BM. In (b), weights are added to the legs of the central square to increase its effective mass density.

The dimensionless parameters determining the qualitative behavior of the UM/BM lattice are the six ratios: $d/a$, $t_1/a$, $t_2/a$, $t_3/a$, $t_x/a$ and $t_y/a$. An ideal effective UM/BM can be obtained if one considers the limit $d/a \to 0$ [26, 29]. The material constants corresponding to aluminum is chosen for the constituent material of the microstructures: $E_s = 69$ GPa, $\rho_s = 2700$ kg/m$^3$ and $v_s = 0.33$. The targeted geometrical parameters used in the following for BM lattice are $l_2 = 5$ mm, $t_3/l_2 = 0.1192$, $t_x/l_2 = 0.358$, $t_y/l_2 = 0.5$ and $l_2/d = 6.5$. The geometrical parameters for UM lattice are determined by Eqns. (18) and (25) together with $t_1/l_1 = 0.1019$ and $t_2/l_1 = 0.06$. Thus, the two lattices have the same lattice constant $a = 15$ mm, hence $d/a \approx 0.05$. Yet smaller ratios of $d/a$ would be desirable, but are now sufficient to illustrate its wave function.

To obtain the static effective elastic constants of the complementary extremal elastic lattice, we fit these elastic constants with the phase velocities in different directions calculated numerically from band structure at low frequency limit for each configuration of solid model (see Refs. [29-30] for detail). Since no resonance effects are considered, the effective mass density is simply calculated by volume average $\rho^{\text{eff}} = \rho V_{\text{solid}}/a^2$, with $V_{\text{solid}}$ being the volume taken by the constituent material in a unit cell. Then, the effective elasticity matrix and effective mass density of the UM in Fig. 3(a) are

$$\mathbf{C}^{\text{UM}} = \begin{bmatrix} 1.6162 & -1.5261 & 0 \\ -1.5261 & 1.6167 & 0 \\ 0 & 0 & 1.6153 \end{bmatrix} \text{GPa}, \quad \rho^{\text{UM}} = 502.37 \text{ kg/m}^3. \tag{26}$$

While, the effective elasticity matrix and effective mass density of the BM in Fig. 3(b) are

$$\mathbf{C}^{\text{BM}} = \begin{bmatrix} 1.5732 & 1.5693 & 0 \\ 1.5693 & 1.5732 & 0 \\ 0 & 0 & 0.0179 \end{bmatrix} \text{GPa}, \quad \rho^{\text{BM}} = 500.44 \text{ kg/m}^3. \tag{27}$$

Therefore, both Eqns. (26) and (27) are nearly isotropic, and their effective elasticity constants and effective mass densities approximately satisfy the relationship of $C_{11}^{\text{UM}} = C_{11}^{\text{BM}}$ and $\rho^{\text{UM}} = \rho^{\text{BM}}$. Fig. 3(b) and (e) show the calculated band structures in the ΓK direction for the solid model plotted in Fig. 3(a) and (b), respectively. In the ΓK direction, the L wave and S wave of the UM are degenerate when $f \leq 35$ kHz, i.e., the velocities of two bulk waves are the same. Besides, the BM supports only L wave when $6 \text{ kHz} \leq f \leq 23 \text{ kHz}$. As a result, when $6 \text{ kHz} \leq f \leq 23 \text{ kHz}$, the wave characteristics of the two type extremal materials are consistent with expectations. The EFCs and polarizations of the two solid models at frequencies of 5 kHz and 15 kHz are shown in Fig. 3(c) and (f), respectively. Again, we obtain a nice consistency between the results of solid models and truss models.

As shown in Fig. 4, a FEA model is designed to examine the mode conversion of the solid extremal material. The geometry of this model can be considered as that obtained by rotating the microstructure in Fig. 2 counterclockwise by $\pi/4$, followed by an appropriately truncation. The total size of the solid model is $297.75 \text{ mm} \times 546.99 \text{ mm}$. Fig. 4(a) present the details at the interface. A modulated Gaussian pulse signal $\cos(2\pi f_c t) \times \exp(-(\pi f_c t/8)^2)$ with a central frequency $f_c = 15$ kHz of upward-propagating L wave is excited from the bottom line source (highlighted in gray in Fig. 4(d)), while the rest boundaries are all set free. The Gaussian pulse has a time duration 0.2 ms, and the Fourier-transformed half amplitude bandwidth is approximately 12 kHz (i.e., a relative bandwidth of 0.8), as shown in Fig. 4(b) and (c), respectively. The calculated displacement amplitude field at $t = 0.3$ and 0.45 ms are shown in panel (d) and (e), respectively, which are conducted by using the Solid Mechanics of COMSOL Multiphysics in the time-domain and normalized by the incident fields. It can be found that the upward-propagating pure L wave excited by the bottom line source can pass through the interface perfectly and transform into an upward-propagating pure S wave, which is consistent with the results of previous analysis.

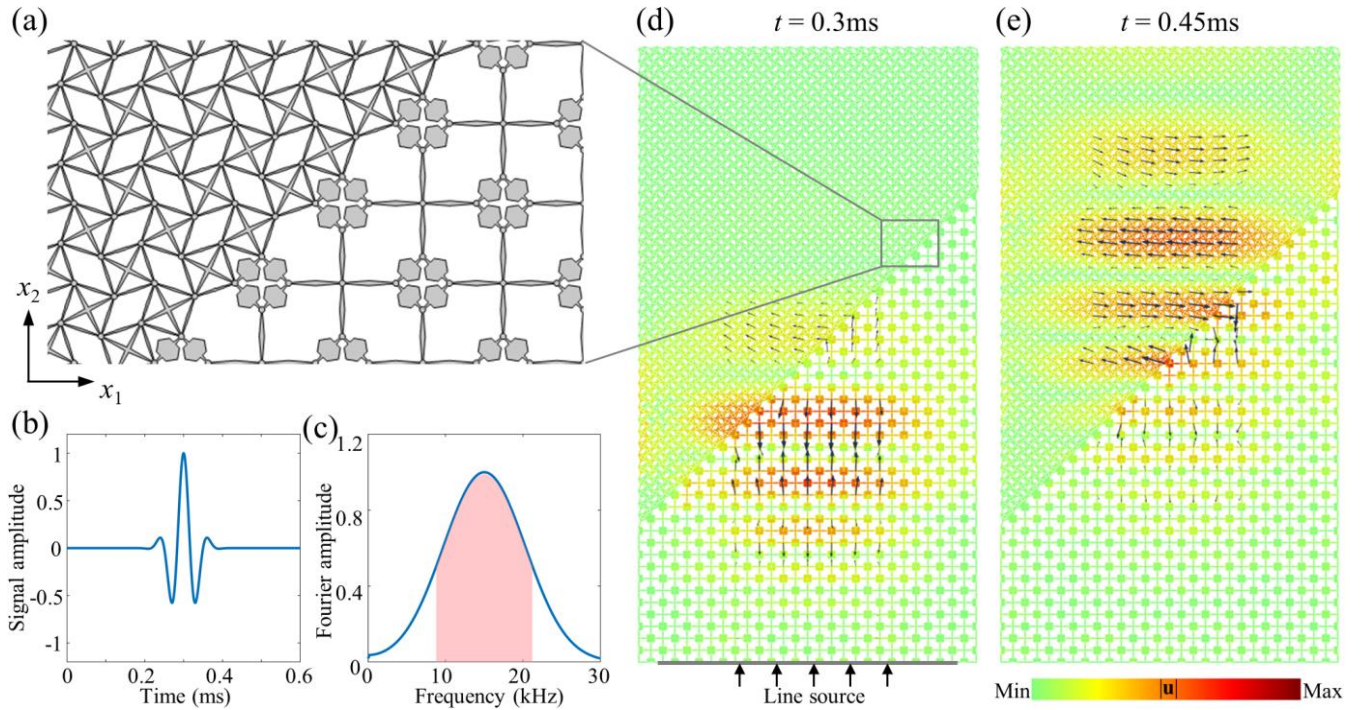

Fig. 4. Verification of the solid model of complementary extremal materials and their wave characteristics. (a) The simulation model consists of isotropic UM/BM solids. (b) Time-domain plot of the Gaussian pulse with a time duration of approximately 0.2 ms used to excite the upward-propagating longitudinal wave. (c) Fourier-transformed amplitude of the Gaussian pulse. The central frequency is 15 kHz and the half amplitude bandwidth is 12 kHz. (d) and (e) are the simulated displacement amplitude at $t = 0.3$ ms and 0.45 ms, respectively. A modulated Gaussian pulse signal of upward-propagating L wave is excited from the bottom line source. The rest boundaries are all set free, which may lead to discrepancies between the displacement vectors at the wave packet boundaries and the expected values. The black arrows represent displacement vectors.

## 4. Flow-permeable underwater acoustic isolator

In 2D case, acoustic fluids (e.g., water) can be treated as isotropic BM, which can only resist hydrostatic stress [13], indicating the wave function demonstrated in Section 3 is naturally expected to be applicable to waterborne sound control. For instance, the BM lattice shown in Fig. 2 has the same material constants as water in long-wavelength limit. Due to only static mechanical properties are utilized without introducing resonant mechanism, it is intrinsically broadband and is of great value in low frequency wave control.

Subsequently, an interesting question arises when considering a block made of isotropic UM with an isosceles right triangle geometry: what propagation characteristics will the underwater acoustic waves exhibit when the block is placed in water? According to the previous analysis, when this block is surrounded by water, the L wave propagating in the $+x_1$ directions in the water would be converted into S wave after passing through the $\pi/4$ inclined surface, as shown in Fig. 5(a). Then, the converted S wave would be normal incident onto the interface between the leg of the block and water, and should be completely reflected due to the decoupling of S and L waves at normal incidence onto a solid-liquid interface [39], i.e., the waterborne sound is blocked. According to the reciprocity principle [40], only S wave propagating in the $-x_1$ directions can successively pass through the leg and hypotenuse of the isosceles right triangle block, and eventually transform into L wave for propagation. Since water only supports the propagation of L wave, waterborne sound traveling in the $-x_1$ directions will be blocked. Following this idea, an array of blocks made of isotropic UM can be designed to forbid waterborne sound propagation.

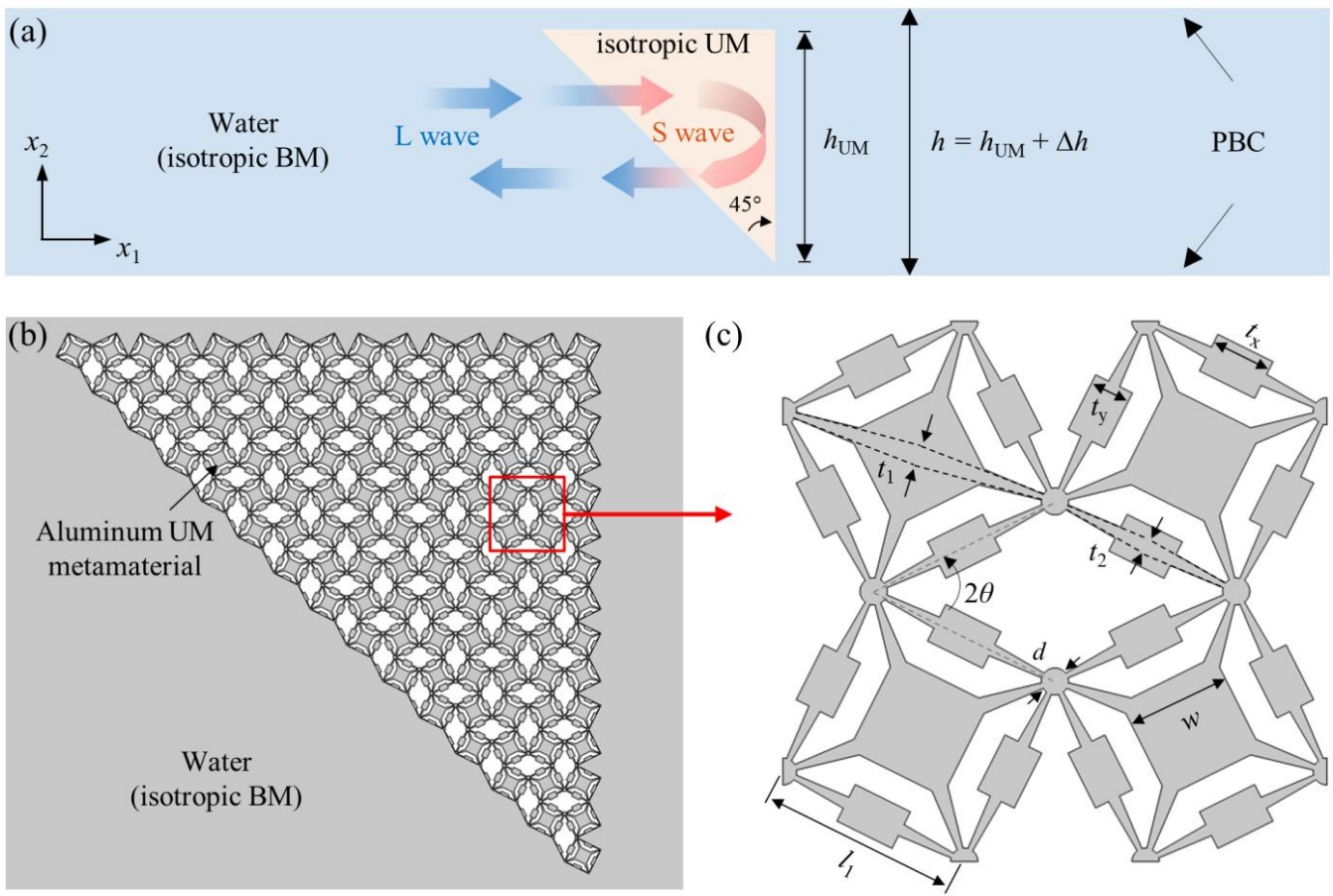

Fig. 5. (a) The principle of flow-permeable underwater acoustic isolator. The light blue background denotes water. Periodic boundary conditions (PBCs) are applied to the top and bottom boundaries of the model to achieve a UM block array. The blue and red arrows represent L and S wave, respectively. (b) The solid model of isotropic UM block with an isosceles right triangle geometry. The length of the leg is $7.5a$ and the triangle block contains 240 small squares in total. Its boundary is modified to connect smoothly to the water. (c) The unit cell of UM. In order to make the UM have the same effective mass density as water, square and rectangular weights are added to its diagonals and sides, respectively.

To reduce computational effort, a solid model of UM block with an isosceles right triangle shape, as shown in Fig. 5(b), is performed for the following wave propagation simulations. It should be noted that in order to make the UM block array forbid the waterborne sound, a certain gap between the blocks is necessary. Then, this underwater acoustic isolator is inherently flow-permeable, and the theoretical water flow rate can be simply evaluated by its open area ratio [41-42]:

$$\delta_\mathrm{V} = \frac{\Delta h}{h} \times 100 \ \%, \tag{28}$$

where $h$ is the height of the background region.

The zoomed view shown in Fig. 5(c) illustrates the detailed configuration of its unit cell, which is obtained by modifying the structure in Fig. 3(a). Aluminum is chosen for the constituent material of the UM. Its geometrical parameters used in the following are $\tan\theta = 0.5$, $l_1 = 12$ mm, $d = 1.61$ mm, $t_1/l_1 = 0.1498$, $t_2/l_1 = 0.08$, $t_\mathrm{x}/t_2 = 3.75$, $t_\mathrm{y}/t_2 = 2.5$ and $w/t_1 = 3.25$. Thus, the lattice constant $a = 32.199$ mm ($d/a \approx 0.05$). Then, the effective elasticity matrix and effective mass density of the UM in Fig. 5(c) are

$$\mathbf{C}^\mathrm{UM} = \begin{bmatrix} 2.1688 & -2.0901 & 0 \\ -2.0901 & 2.1692 & 0 \\ 0 & 0 & 2.1121 \end{bmatrix} \text{GPa}, \ \rho^\mathrm{UM} = 1058.91 \text{ kg/m}^3. \tag{29}$$

As a result, the magnitudes of effective elasticity constants and effective mass density are approximately the same as those of water. Substituting Eq. (29) back into Eq. (3) yields the bulk wave velocities along the ΓK direction:

$$c_\mathrm{L}^{[110]} \approx 1431.14 \text{ m/s}, \ c_\mathrm{T}^{[110]} \approx 1412.31 \text{ m/s}, \tag{30}$$

indicating that the UM achieves impedance matching of the L wave with water $\rho^\mathrm{UM} c_\mathrm{L}^{[110]} / \rho^\mathrm{W} c_\mathrm{W} \approx 1.01$, where $\rho^\mathrm{W}$ and $c_\mathrm{W}$ respectively denotes the mass density and sound velocity of water.

The simulations are performed using coupled acoustic-solid module in COMSOL Multiphysics in the frequency domain. An acoustic pressure field of the form $p_\mathrm{IN} = \exp(-\mathrm{i}k_1 x)$ ($k_1 = 2\pi f/c_\mathrm{W}$ is the wavenumber component along $x_1$-axis) is imposed to generate waterborne sound with an incident amplitude of 1 Pa on the UM block. The PMLs are added on the left and right sides to eliminate the influence of reflections. The sound-transmission loss $L_\mathrm{st}$ of the isolator can be then evaluated by

$$L_\mathrm{st} = -20\log_{10}\left|p_\mathrm{T}/p_\mathrm{IN}\right|, \tag{31}$$

in which, $p_\mathrm{T}$ denotes the amplitude of the transmitted wave.

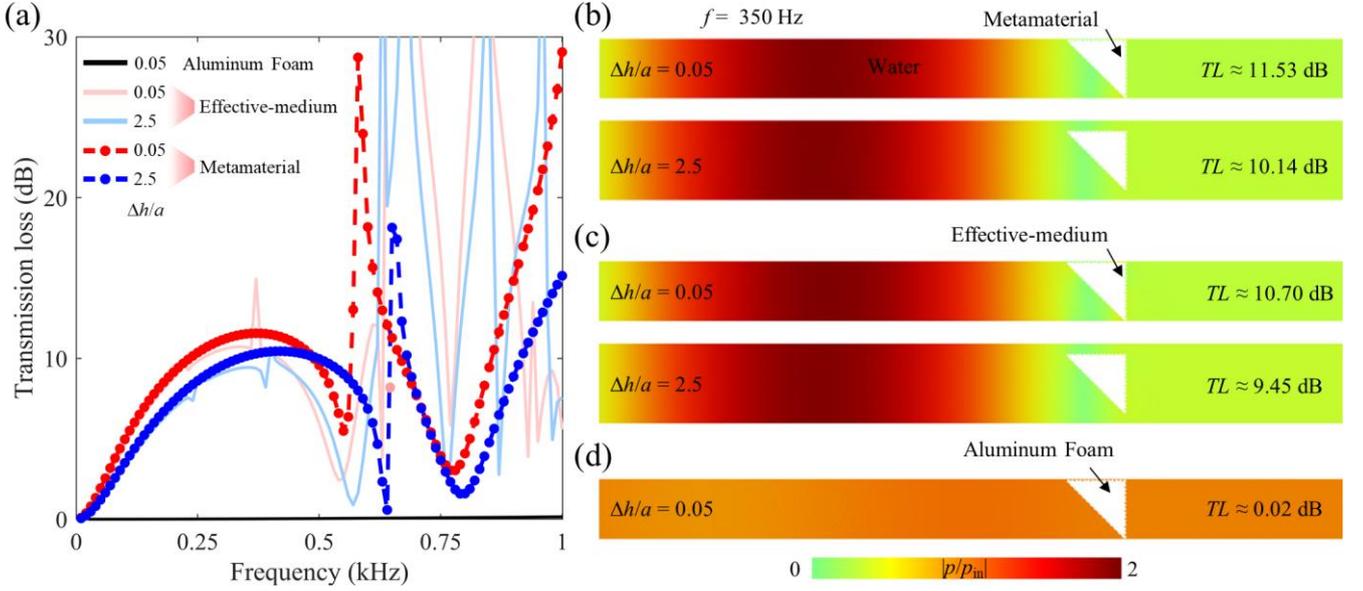

Fig. 6. Simulation results for illustrating the functionality of the UM-based waterborne sound isolator. (a) the sound-transmission loss $L_{st}$ versus the incident wave frequency for $\delta_V = 0.67\%$ and $25\%$ (or equivalently $\Delta h/a = 0.05$ and $2.5$), respectively. (b)-(d) The simulated amplitudes of pressure fields at $\Delta h/a = 0.05$ and $2.5$ at $f = 350$ Hz for the middle triangular region occupied by the Al-based UM, effective-medium and aluminum foam, respectively.

Figure 6(a) shows the computed sound-transmission loss $L_{st}$ versus the incident wave frequency for different $\delta_V$, where $\delta_V = 0.67\%$ and $25\%$ (or equivalently $\Delta h/a = 0.05$ and $2.5$), respectively. When $\delta_V \leq 25\%$ and $f \leq 1$ kHz, the variation in $L_{st}$ is insignificant, and the results obtained from discreate and the continuum models show good agreement. When $\Delta h/a = 0.05$ and $f = 350$ Hz, the discreate model exhibits excellent sound insulation performance ($L_{st} \approx 11.53$ dB), and the simulated amplitude of the pressure fields is shown in Figure 6(b). The normalized wavelength of the incident wave is $\bar{\lambda} = c_W/(h_{UM}f) \approx 17.75$. In addition, in panel (b), when $\Delta h/a = 2.5$, although the water flow rate remains high, the incident wave is still largely reflected by the UM array, resulting in a weak transmitted wave and a considerable sound-transmission loss $L_{st} \approx 10.14$ dB. For comparison, the region occupied by the UM is entirely replaced with its counterpart and aluminum foam ($E_{sf} = (\rho_{sf}/\rho_s)^2 E_s$, $\rho_{sf} = \rho^{UM}$ and $\nu_{sf} = 0.3$ [43]), with the corresponding results shown in Fig. 6(c) and (d), respectively. Again, we obtain a nice consistency between the results of discreate and continuum models. Furthermore, in Fig. 6(d), the amplitude of transmitted wave is markedly enhanced, while only a negligible portion of the incident wave is reflected, indicating a very limited capacity to block waterborne sound ($L_{st} = 0.02$ dB).

## 5. Conclusions

We have proposed a concept of complementary extremal materials and examined wave characteristics at their interface. In 2D case, the two bulk modes in isotropic UM are degenerate and support any polarization, which is significantly different from conventional Cauchy materials. When L wave in an isotropic BM is incident at an angle of $\pi/4$ onto the interface with certain isotropic UM

(complementary to isotropic BM), it will be completely converted into S wave, i.e., a perfect mode conversion, which can be also regarded as an elastic wave diode. Since only static mechanical properties are utilized without introducing resonant mechanism, the mechanism is intrinsically broadband and being of great value for subwavelength elastic wave manipulation. The corresponding truss and solid models are also designed, and the effective material constants of truss models are given analytically. The function of broadband mode conversion is demonstrated by both homogenized and discrete models.

As an application, we have numerically demonstrated a flow-permeable waterborne sound isolator composed of an array of isotropic UM blocks with isosceles right triangles shapes. The simulated results show the effectiveness of our proposed mechanism to block low frequency waterborne sound while allowing a high ratio of the water flow to pass, e.g., the Al-based UM blocks array achieves $L_{st} \approx$ 10.14 dB when $\delta_V = 25\%$ and $f = 350$ Hz. This study paves the way for exploring the exotic wave properties of complementary extremal materials, and opens a new route to control low frequency elastic and underwater acoustic waves.

## 6. Acknowledgments

We acknowledge the support from National Natural Science Foundation of China (Grant No. 12532006), Zhuhai Basic and Applied Basic Research Foundation (Grant No. 2320004002694).

## Appendix

Firstly, consider the anisotropic case of UM with Eq. (5), whose dispersion relation and polarization are expressed as Eqns. (8) and (9), respectively. When $\alpha = +1$, the EFC of the first mode has four openings along the following directions:

$$n_1^2 = \frac{\sqrt{C_{22}}}{\sqrt{C_{11}} + \sqrt{C_{22}}}, \quad n_2^2 = \frac{\sqrt{C_{11}}}{\sqrt{C_{11}} + \sqrt{C_{22}}}, \tag{32}$$

which is obtained by performing some math operation on Eq. (8). The EFC of the second mode is closed and shaped like a rounded square. However, $\alpha = -1$ leading to two closed EFCs. For example, let $C_{11} = 1$, $C_{22} = 2$, $C_{33} = 3$ and $\rho = 1$, the EFCs and polarizations of the UM for $\alpha = +1$ and $\alpha = -1$ are shown in Fig. A1 (a) and (b), respectively, which are in good agreement with the above analysis. The dispersion relation and bulk wave polarizations of the BM complementary to this type UM are expressed as Eqns. (10) and (11), respectively, indicating that the sign of $\alpha$ affects only the polarizations, while having no influence on the shape of the EFC.

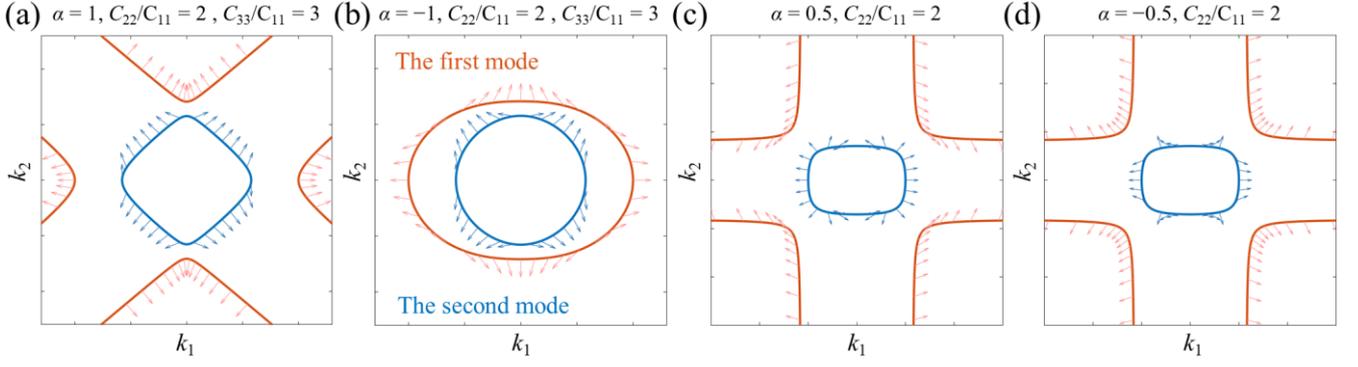

Fig. A1. EFCs and polarizations. (a)-(b) and (c)-(d) are UM whose elasticity matrix have the form of Eq. (5) and Eq. (6), respectively. The red and blue solid lines represent the first and second modes, respectively. The arrows denote polarizations.

Then, for a UM characterized by the elasticity matrix of Eq. (6), the shear stress constitutes its soft mode. The $\alpha > 0$ and $\alpha < 0$ indicate UM possessing positive and negative Poisson's ratio, respectively. In this case, any BM with elasticity matrix form as $\mathbf{C} = \text{diag}(0, 0, C_{33})$ is complementary to such UM, i.e., it is not unique. Substituting the UM elasticity matrix of Eq. (6) into Eq. (4) yields the dispersion relation

$$\omega_1^2 = \frac{1}{2\rho}\left(C_{11}k_1^2 + C_{22}k_2^2 - \sqrt{4\alpha^2 C_{11}C_{22}k_1^2 k_2^2 + \left(C_{11}k_1^2 - C_{22}k_2^2\right)^2}\right),$$
$$\omega_2^2 = \frac{1}{2\rho}\left(C_{11}k_1^2 + C_{22}k_2^2 + \sqrt{4\alpha^2 C_{11}C_{22}k_1^2 k_2^2 + \left(C_{11}k_1^2 - C_{22}k_2^2\right)^2}\right),$$
(33)

The two bulk wave modes corresponding to Eq. (33) are as follows

$$\hat{\mathbf{u}}_1 = \begin{bmatrix} C_{11}n_1^2 - C_{22}n_2^2 - \sqrt{4\alpha^2 C_{11}C_{22}n_1^2 n_2^2 + \left(C_{11}n_1^2 - C_{22}n_2^2\right)^2} \\ 2\alpha\sqrt{C_{11}C_{22}}\,n_1 n_2 \end{bmatrix},$$
$$\hat{\mathbf{u}}_2 = \begin{bmatrix} C_{11}n_1^2 - C_{22}n_2^2 + \sqrt{4\alpha^2 C_{11}C_{22}n_1^2 n_2^2 + \left(C_{11}n_1^2 - C_{22}n_2^2\right)^2} \\ 2\alpha\sqrt{C_{11}C_{22}}\,n_1 n_2 \end{bmatrix},$$
(34)

which indicating that varying the $C_{11}$ and $C_{22}$ can lead to a rich diversity in the polarization and EFCs. However, the sign of $\alpha$ affects only the polarizations but not change the shape of the EFCs, see also Fig. A1(c) and (d). For example, let $C_{22} = 2C_{11} = 2$ and $\rho = 1$, the EFCs and polarizations of the UM with $\alpha = 0.5$ and $-0.5$ are shown in Fig. A1 (c) and (d), respectively. The EFC of the first mode has four openings along the $x_1$- and $x_2$-axis, whereas that of the second mode is closed and shaped like a rounded rectangle.

Furthermore, substituting the elasticity matrix $\mathbf{C} = \text{diag}(0, 0, C_{33})$ (i.e., the BM complemented to the UM of Eq. (6)) into Eq. (4) yields the dispersion relation

$$\omega_1^2 = 0, \quad \omega_2^2 = k^2 C_{33} / \rho,$$
(35)

and the two bulk wave modes corresponding to Eq. (35) are as follows

$$\hat{\mathbf{u}}_1 = \begin{bmatrix} -n_1 \\ n_2 \end{bmatrix}, \hat{\mathbf{u}}_2 = \begin{bmatrix} n_2 \\ n_1 \end{bmatrix}. \tag{36}$$

Therefore, the BM possess only one non-vanishing bulk wave, and the corresponding EFC is a closed circle. Besides, the polarization $\hat{\mathbf{u}}_2$ and wave propagation $\mathbf{n}$ are always symmetric with respect to the diagonal of the $(x_1, x_2)$-plane.